%% file: prd.tex
\documentclass[aps,prd,twocolumn,showpacs,superscriptaddress,groupedaddress]{revtex4}
\usepackage{graphicx}
\usepackage{dcolumn}
\usepackage{bm}
\usepackage{amssymb}
\usepackage{amsmath}
\usepackage{xspace}
\usepackage{color}
\usepackage{enumerate}
\usepackage{afterpage}
\newcommand{\ppbar}{$p\bar{p}$~}

\newcommand{\pt}{${p}_{T}$\xspace}

\newcommand{\Ptg}{p_{T}^{\gamma}}
\newcommand{\Ptj}{p_{T}^\text{jet}}

\newcommand{\ve}{\varepsilon}

\newcommand{\la}{\langle}
\newcommand{\ra}{\rangle}

\newcommand{\gt}{\!>\!}

\newcommand{\sigmaeff}{$\sigma_{\rm eff}$}
\newcommand{\sigmaeffhf}{$\sigma_{\rm eff}^{\rm HF}$}

\newcommand{\gpj}{$\gamma+{\rm jet}$\xspace}
\newcommand{\gpTHRj}{$\gamma+{\rm 3~jet}$\xspace}
\newcommand{\gpHFjj}{$\gamma+{b/c~\rm jet+2~jet}$\xspace}

\newcommand{\DPhi}{$\Delta\phi(\gamma,\rm jet1)$\xspace}
\newcommand{\dPhi}{$\Delta S$\xspace}

\newcommand{\ndp}{$N_{\mathrm{DP}}$\xspace}
\newcommand{\ndi}{$N_{\mathrm{DI}}$\xspace}
\newcommand{\fdp}{$f_{\mathrm{DP}}$\xspace}
\newcommand{\fdi}{$f_{\mathrm{DI}}$\xspace}

\newcommand{\chindf}{$\chi^2/ndf$\xspace}

\newcommand{\epsdi}{\varepsilon_{\rm DI}}
\newcommand{\epsdp}{\varepsilon_{\rm DP}}

\newcommand{\onevtx}{$\textsc{1vtx}$\xspace}
\newcommand{\twovtx}{$\textsc{2vtx}$\xspace}

\newcommand{\mixdp}{$\textsc{mixdp}$\xspace}
\newcommand{\mixdi}{$\textsc{mixdi}$\xspace}
\newcommand{\mcdp}{$\textsc{mcdp}$\xspace}
\newcommand{\mcdi}{$\textsc{mcdi}$\xspace}
\newcommand{\sponevtx}{$\textsc{sp1vtx}$\xspace}
\newcommand{\sptwovtx}{$\textsc{sp2vtx}$\xspace}
\newcommand{\pythia}{$\textsc{pythia}$\xspace}
\newcommand{\sherpa}{$\textsc{sherpa}$\xspace}

\newcommand{\GeV}{\ensuremath{\text{GeV}}\xspace}

\begin{document}

\hspace{5.2in} \mbox{FERMILAB-PUB-14-018-E}
\title{\boldmath 
Double parton interactions in \gpTHRj and \gpHFjj events in $p\bar{p}$ collisions
at $\sqrt{s}=1.96$~TeV}
\input author_list.tex

\date{February 6, 2014}
%\date{\today}

\begin{abstract}
     We determine the fraction of events with double parton (DP)
     scattering in a single $p \bar{p}$ collision at $\sqrt{s}=1.96$ TeV
     in samples of \gpTHRj and \gpHFjj
     events collected with the D0 detector 
     and corresponding to an integrated luminosity of about 8.7~fb$^{-1}$.
     The DP fractions and effective cross sections ($\sigma_{\rm eff}$) 
     are measured for both event samples
     using the same kinematic selections. 
     The measured DP fractions range from $0.21$ to $0.17$, with effective cross sections 
     in the \gpTHRj and \gpHFjj samples of 
     $\sigma_{\rm eff}^{\rm incl} = 12.7 \pm 0.2\thinspace(\rm stat) \pm 1.3\thinspace(\rm syst)$~mb and
     $\sigma_{\rm eff}^{\rm HF} = 14.6 \pm 0.6\thinspace(\rm stat) \pm 3.2\thinspace(\rm syst)$~mb, respectively.

\end{abstract}
\maketitle

%%%%%%%%%%%%%%%%%%%%%%%%%%%%%%%%%%%%%%%%%%%%%%%%%%%%%%%%%%%%%%%%%%%%%%%%%%%%%%%%%%%%%%%
%%% Main text starts here

%\setpagewiselinenumbers
%\linenumbers

\section{Introduction}
\label{Sec:Intro}

The study of deep inelastic hadron-hadron collisions is one of the 
main sources of knowledge about hadronic structure. 
We describe such a collision as the
process in which a single parton (quark or gluon) from one nucleon undergoes a hard scattering
off a single parton from the other nucleon. 
The other ``spectator'' partons, which do not take part in this hard $2 \to 2$
parton collision, contribute to the so-called ``underlying event.''
However, the probability of other partons in each nucleon to also undergo a hard scattering
is not zero. 
The rate of multiple parton interactions (MPI) in $p \bar{p}$ collisions is directly
related to the transverse spatial distribution of partons within the proton, and has
been the subject of extensive theoretical
studies (see e.g., \cite{Landsh, TH1, TH2, TH3, Threl, Threl2, Flen, Sjost, Snigir, Strikman}).

Relevant measurements have been performed by the AFS~\cite{AFS}, UA2~\cite{UA2},
CDF~\cite{CDF93,CDF97}, D0~\cite{D02010,D02011}, ATLAS~\cite{ATLAS2013}, and CMS~\cite{CMS2013}
collaborations.
The first three measurements are based on 
samples of events having a 4-jet final state,
while the CDF and D0 measurements in Refs.~\cite{CDF97,D02010,D02011} 
use \gpTHRj events produced by double parton (DP) scattering with \gpj and dijet final states.
The \gpj production originates mainly via 
quark-gluon scattering in a Compton-like process, 
$qg \to q \gamma$, and an annihilation process, $q\bar{q} \to g \gamma$. 
As was shown experimentally
in Refs.~\cite{CDF93,CDF97,D02010,D02011} and theoretically described in Ref.~\cite{Han}, 
the use of \gpTHRj events leads to a greater sensitivity 
to the DP fraction as compared to 4-jet events 
mainly because of the better energy and angular resolutions for photon 
as compared with jets.

The total DP cross section $\sigma_{\rm DP}$ for the events
caused by two parton scatterings with \gpj and dijet final states
is defined as~\cite{D02010}
\begin{eqnarray}
\sigma_{DP} =
 \frac{\sigma^{\gamma j} \sigma^{jj}}{\sigma_{\rm eff}}~.
\label{eq:sigmaeff_main}
\end{eqnarray}
Here, $\sigma^{\gamma j}$ ($\sigma^{jj}$) is the total \gpj (dijet) production
cross section.
The parameter $\sigma_{\rm eff}$ in Eq.~\ref{eq:sigmaeff_main} is related to the distance
in the transverse plane between partons in the
nucleon \cite{TH1, TH2, Threl, AFS, UA2, CDF93, CDF97, D02010}:
\begin{eqnarray}
\sigma_{\rm eff} = \Bigl[\int (F(\beta))^2 d^2 \beta \Bigr]^{-1},
\label{eq:sigmaeff_spatial}
\end{eqnarray}
where $F(\beta)=\int \rho(r) \rho(r-\beta)d^2r$ is the overlap function
between the parton spatial distributions $\rho(r)$ 
in the nucleons colliding with impact parameter $\beta$ (for example, see \cite{Threl,Threl2,Flen}).
Here $r$ is a distance from the center of the nucleon in transverse plane. 
The overlap function is normalized to unity, $\int F(\beta) d^2 \beta =1$.
In case of a Gaussian spatial density $\rho(r)$, 
the overlap function $F(\beta)=(4\pi a^2)^{-1} \exp(-\beta^2/2a^2)$, and thus $\sigma_{\rm eff} = 8\pi a^2$,
where $a$ is the Gaussian width~\cite{Flen, D02010}.
The overlap function characterizes the transverse area occupied by the interacting partons.
The larger the overlap (i.e., smaller $\beta$), the more probable it is to have one or more
hard parton interactions in the colliding nucleons.

Table~\ref{tab:sigeff_world} summarizes the currently available measurements of the
value of $\sigma_{\rm eff}$.  Within uncertainties, existing measurements 
of $\sigma_{\rm eff}$  for final states with jets and photons or $W$ bosons are consistent.
They are more precise than those with 4-jet final state.
The dependence of $\sigma_{\rm eff}$ on $\sqrt{s}$ is expected to be small \cite{Threl}.

\begin{figure}[h]
\vspace*{0mm}
\hspace*{0mm} \includegraphics[scale=0.37]{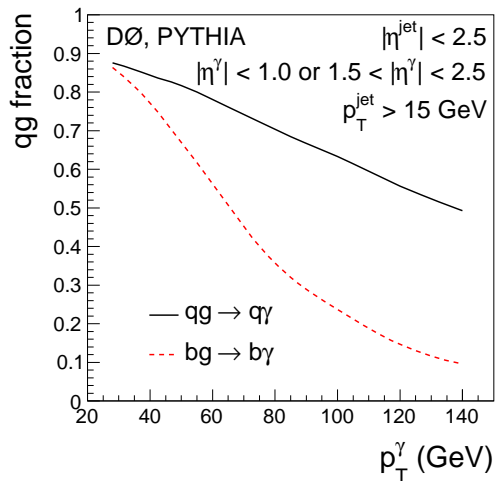}
\vspace*{-2mm}
\caption{(color online)
Fractional contribution of the Compton-like $qg\to q \gamma$ ($q$ is any quark type) 
and $bg\to b \gamma$ subprocesses to the associated production of 
inclusive \gpj and $\gamma+b$-jet final states as a function of $\Ptg$.
}
\label{fig:qg_frac}
\vspace*{5mm}
\end{figure}

In this paper we present the first measurement of the DP rates and $\sigma_{\rm eff}$ involving heavy flavor
leading jet using the \gpHFjj final state and compare this measurement 
to the results obtained with \gpTHRj events.
The $\gamma+b/c$-jet production is mainly caused by
$b(c)g \to b(c) \gamma$ and $q\bar{q} \to g \gamma$ with $g \to Q\bar{Q}$, where $Q = b(c)$~\cite{gamma_b}.
\begin{table*}[htpb]
\vskip1mm
\begin{center}
\caption{Summary of the results, experimental parameters, and event selection criteria for the double parton analyses 
performed by the AFS, UA2, CDF, D0, ATLAS, and CMS Collaborations 
(no uncertainties are available for the AFS result).}
\label{tab:sigeff_world}
\vskip 1mm
\begin{tabular}{|c||cclll|} \hline
$  $ & $ \sqrt{s}$ (GeV) & final state & ~~$p_{T}^{\rm cut}$ (GeV)~~ & $\eta $ range & $\sigma_{\rm eff}$ \\\hline
 AFS~\cite{AFS} & $ 63 $  &  4~jets       & $\Ptj>4$ & $|\eta^{\rm jet}|<1$  &  $\approx 5 $ mb \\\hline  
 UA2~\cite{UA2} & $ 630 $  & 4~jets       & $\Ptj>15$ & $|\eta^{\rm jet}|<2$  & $ > 8.3$ mb (95\% C.L.) \\\hline
 CDF~\cite{CDF93} & $ 1800 $ &  4~jets       & $\Ptj>25$ & $|\eta^{\rm jet}|<3.5$ & $ 12.1_{-5.4}^{+10.7}$ mb \\\hline
 CDF~\cite{CDF97} & $ 1800 $ & $\gamma + 3$~jets  & $\Ptj>6$  & $|\eta^{\rm jet}|<3.5$ & $14.5 \pm 1.7\thinspace({\rm stat})~{}_{-2.3}^{+1.7}\thinspace({\rm syst})$ mb  \\
     &      &           & $\Ptg>16$ & $|\eta^{\gamma}|<0.9$ &                              \\\hline
 D0~\cite{D02010} & $ 1960 $ & $\gamma + 3$~jets   & $60<\Ptg<80$&$|\eta^{\gamma}|<1.0$ & $16.4 \pm 0.3\thinspace({\rm stat}) \pm 2.3\thinspace({\rm syst})$ mb \\
      &      &           &  $\Ptj>15$     &$1.5<|\eta^{\gamma}|<2.5$ &                     \\\hline
 ATLAS~\cite{ATLAS2013} & $ 7000 $ & $W + 2$~jets    & $\Ptj>20$& $|\eta^{\rm jet}|<2.8$ & $15 \pm 3\thinspace({\rm stat})~{}_{-3}^{+5}\thinspace({\rm syst})$ mb\\\hline
   CMS~\cite{CMS2013} & $ 7000 $ & $W + 2$~jets    & $\Ptj>20$& $|\eta^{\rm jet}|<2.0$ & $20.7 \pm 0.8\thinspace({\rm stat}) \pm 6.6\thinspace({\rm syst})$ mb\\\hline
\end{tabular}
\end{center}
\vskip 2mm
\end{table*}
Figure~\ref{fig:qg_frac} shows the fractions of $gq$ and $gb$ subprocesses
in events with \gpj and $\gamma+b$-jet final states,
calculated using default {\sc pythia 6.4} \cite{PYT} settings
and the {\sc cteq} 6.1L parton distribution function~\cite{CTEQ}.
At $\Ptg\approx 30$ GeV, Compton-like scattering dominates over 
the annihilation process, contributing about 85\%--88\% of events.
Since the initial quarks in the Compton-like scattering for inclusive \gpj and $\gamma+b/c$-jet
production are typically light ($\approx 92\%$, according to the estimates
done with {\sc pythia}) and $b/c$ quarks, respectively,  
the difference between effective cross sections measured in the two processes
should be sensitive to 
difference between light quark and heavy quark transverse spatial distributions
(see Eq.~\ref{eq:sigmaeff_spatial}).

The outline of the paper is as follows.
Section \ref{sec:Method} briefly describes the technique for extracting 
the effective cross section $\sigma_{\rm eff}$. 
Section \ref{sec:ObjectID} includes the 
description of the D0 detector and the data and Monte Carlo simulation (MC) samples used in the measurement.
Section \ref{sec:Models} presents signal and background models. Section \ref{sec:Vars}
describes the discriminating variable used to determine the DP fractions.
The DP fractions are estimated in Section \ref{sec:Frac}.
Section \ref{sec:Eff} describes
the determination of other parameters needed to calculate  $\sigma_{\rm eff}$.
In Section~\ref{sec:Sigma_eff}, we calculate the effective cross section
$\sigma_{\rm eff}$ for \gpTHRj and \gpHFjj events,
and discuss the effects related to 
parton distribution functions (PDF) in Section~\ref{sec:PDF}.
The results are summarized in Section~\ref{sec:summary}.

\section{Technique for extracting $\sigma_{\rm eff}$ from data}
\label{sec:Method}

    To extract $\sigma_{\rm eff}$, we use the same
    technique as in earlier measurements~\cite{CDF97,D02010}, 
    which requires only quantities
    determined from data, minimizing the impact of theoretical assumptions. 
    We avoid using theoretical predictions of the \gpj and dijet cross sections
    by comparing the number of \gpTHRj events produced in DP interactions
    in single \ppbar collisions to the number of \gpTHRj events produced in two distinct
    hard parton interactions occurring in two separate \ppbar collisions in the same beam crossing.
    The latter class of events is referred to as double interaction (DI) events.  
    Assuming uncorrelated parton scatterings in the DP process~\cite{Landsh,TH1,TH2,TH3,Threl},
    DP and DI events should be kinematically identical,
    and only differ by the presence of one (two) \ppbar collision vertex in the case of DP (DI) events. 
    This assumption has been tested in Ref.~\cite{D02010} and is discussed further in Section~\ref{sec:Sigma_eff}. 
    Both DP and DI interactions provide a source of events with two instances of parton scattering.  
    It is necessary to measure both DP and DI rates to extract $\sigma_{\rm eff}$.
    Background processes include single hard interactions producing similar final states 
    with or without the presence of additional soft \ppbar interactions.

   As was shown in Ref.~\cite{D02010}, 
   the number of DI events with the final topology of interest, $N_{\rm DI}$, 
   can be obtained from the probability for a DI event, 
   $P_{\rm DI} = 2(\sigma^{\gamma j} / \sigma_{\rm hard})(\sigma^{jj} / \sigma_{\rm hard})$, 
   in a \ppbar beam crossing with two
   hard collisions.
   Here $\sigma_{\rm hard}$ is the total hard \ppbar interaction cross section.
   This probability should be corrected for the 
   combination of the acceptance (geometric and kinematic) and
   selection efficiency ($\epsilon_{\rm DI}$), 
   the two-vertex event selection efficiency ($\ve_{\rm 2vtx}$), 
   and the number of beam crossings with two hard collisions ($N_{\rm 2coll}$):
   \begin{eqnarray}
   N_{\rm DI} = P_{\rm DI}N_{\rm 2coll}\epsilon_{\rm DI}\ve_{\rm 2vtx}.
   \label{eq:N_DI}
   \end{eqnarray}

   Analogously to $N_{\rm DI}$, the number of DP events, $N_{\rm DP}$, can be expressed from 
   the probability for a DP event, 
   $P_{\rm DP} = (\sigma^{\gamma j} / \sigma_{\rm hard})(\sigma^{jj} / \sigma_{\rm eff})$, 
   in a \ppbar beam crossing with one
   hard collision.
   Similarly to the DI events, this probability is corrected for the 
   combination of the acceptance (geometric and kinematic) and
   selection efficiency ($\epsilon_{\rm DP}$), 
   the single-vertex event selection efficiency ($\ve_{\rm 1vtx}$),
   and the number of beam crossings with one hard collision ($N_{\rm 1coll}$):
   \begin{eqnarray}
   N_{\rm DP} = P_{\rm DP}N_{\rm 1coll}\epsilon_{\rm DP}\ve_{\rm 1vtx}.
   \label{eq:N_DI}
   \end{eqnarray}

   The ratio of the number of DP to DI events, $N_{\rm DP}/N_{\rm DI}$, 
   allows us to obtain the expression for $\sigma_{\rm eff}$~\cite{CDF97,D02010}: 
   \begin{eqnarray}   
   \sigma_{\rm eff} = \frac{N_{\rm DI}}{N_{\rm DP}} \frac{\ve_{\rm DP}}{\ve_{\rm DI}} R_{\rm c}\sigma_{\rm hard},
   \label{eq:sig_eff}
   \end{eqnarray}
   where the factor $R_c \equiv (1/2) (N_{\rm 1coll} /N_{\rm 2coll}) (\ve_{\rm 1vtx} / \ve_{\rm 2vtx})$.
   The cross sections $\sigma^{\gamma j}$ and $\sigma^{jj}$ do not appear in this equation, 
   and all efficiencies for DP and DI events enter only as ratios,
   resulting in a reduction of the correlated systematic uncertainties.

   The background to DP events are single parton (SP) scatterings with 
   the radiation of at least two hard gluons
   in the initial or final state, $qg \to q\gamma gg$, $q\bar{q} \to g\gamma gg$,
   which leads to the same \gpTHRj signature.
   The fraction of DP events is determined using a variable sensitive to the
   kinematic configurations of the two independent scatterings of parton pairs.

   The largest background to DI events
   is two-vertex SP events with one hard \gpTHRj interaction occurring in one \ppbar collision
   and an additional soft interaction (i.e., having no reconstructed jets) 
   occurring at the other \ppbar vertex.

\section{D0 detector and data samples}
\label{sec:ObjectID}

The D0 detector is described in detail in Refs.~\cite{d0det,l1cal,l0}. Photon candidates are
identified as isolated clusters of 
energy depositions in one of three uranium and liquid argon sampling calorimeters. 
The central calorimeter covers the pseudorapidity~\cite{d0_coordinate} range $|\eta_{\rm det}|<1.1$,
and the two end calorimeters cover up to $|\eta_{\rm det}| \approx 4.2$.
In addition, the plastic scintillator intercryostat detector covers
the region $1.1<|\eta_{\rm det}|<1.4$.
The electromagnetic (EM) section of the calorimeter is segmented longitudinally into 
four layers and transversely into cells in pseudorapidity and azimuthal angle 
$\Delta\eta_{\rm det}\times\Delta\phi_{\rm det}=0.1 \times 0.1$ 
($0.05 \times 0.05$ in the third layer of the EM calorimeter).
The hadronic portion of the calorimeter is located behind the EM section.
The calorimeter surrounds 
a tracking system consisting of a silicon microstrip tracking (SMT) detector and
scintillating fiber tracker, both located within a 2~T solenoidal magnetic field.
The solenoid magnet is surrounded by the central preshower (CPS)
detector located immediately before the calorimeter. The CPS consists of
approximately one radiation length of lead absorber surrounded by three layers of scintillating strips.

The current measurement is based on 8.7~fb$^{-1}$ of data collected
after the D0 detector upgrade in 2006~\cite{l0}, 
while the previous measurements~\cite{D02010, D02011} were made using data collected before this upgrade.
 
The events used in this analysis pass triggers designed to identify
high-\pt clusters in the EM calorimeter with loose
shower shape requirements for photons.
These triggers have $\approx 96\%$ efficiency at $\Ptg \approx 30$~GeV and are $100\%$ efficient
for $\Ptg \gt 35$~GeV.

To select photon candidates in our data samples,
we use the following criteria~\cite{gamjet_tgj,EMID_NIM}:
EM objects are reconstructed using a simple cone algorithm with
a cone size of $\Delta{\cal R}=\sqrt{(\Delta\eta)^2 + (\Delta\phi)^2}=0.2$.
Regions with poor photon identification capability and degraded $\Ptg$ resolution
at the boundaries between calorimeter modules and between the central and endcap
calorimeters are excluded from analysis.
Each photon candidate is required to deposit more than 96\% of the detected energy
in the EM section of the calorimeter 
and to be isolated in the angular region between
$\Delta{\cal R}=0.2$ and $\Delta{\cal R}=0.4$ around the 
center of the cluster:
$(E^{\rm iso}_{\rm tot}-E^{\rm iso}_{\rm core})/E^{\rm iso}_{\rm core} < 0.07$, where $E^{\rm iso}_{\rm tot}$ 
is the total (EM+hadronic) tower energy in the ($\eta,\phi$) cone of radius $\Delta{\cal R}=0.4$
and $E^{\rm iso}_{\rm core}$ is EM energy within a radius of $\Delta{\cal R}=0.2$.
Candidate EM clusters that match to a reconstructed track are excluded from the analysis. 
We also require the energy-weighted EM cluster width in the finely-segmented third EM layer
to be consistent with that expected for a photon-initiated electromagnetic shower.
In addition to the calorimeter isolation cut, we also apply a track isolation cut,
requiring the scalar sum of track transverse
momenta in an annulus $0.05 \leq \Delta{\cal R} \leq 0.4$ to be less than 1.5~GeV.

Jets are reconstructed using 
an iterative midpoint cone algorithm~\cite{JetAlgo}
with a cone size of $0.5$. Jets must satisfy
quality criteria that suppress background from leptons, photons, and
detector noise effects.
Jet transverse momenta are corrected to the particle level~\cite{JES_NIM}.

To reject background from cosmic rays and $W\to e\nu$ decay~\cite{gamjet_tgj},
the missing transverse momentum in the event is required to be less than $0.7p_{T}^{\gamma}$.
All photon-jet pairs must be separated by $\Delta{\cal{R}}>0.7$ 
and all jet-jet pairs must be separated by $\Delta{\cal{R}}>1.0$.
Each event must contain at least one photon in the pseudorapidity
region $|\eta^{\gamma}|<1.0$ or $1.5<|\eta^{\gamma}|<2.5$ and at least three jets with $|\eta^{\rm jet}|<2.5$.
The jet with the highest \pt is termed the ``leading jet'' or first jet,
and the jets with the second and third highest \pt are denoted as 
the second and third jets in the following.
Events are selected with photon transverse momentum 
$p^{\gamma}_{T}>26$~GeV, leading jet $p_T^{\rm jet}>15$~GeV, while
the next-to-leading (second) and third jets must have $15<p_T^{\rm jet}<35$~GeV.
The upper limit on the \pt of the second and third jets increases the
fraction of DP events in the sample~\cite{D02010}.

To select the sample of \gpHFjj candidate events,
the leading jet is required to have at least two associated tracks with $p_T>0.5$~\GeV
and each track must have at least one hit in the SMT detector.
At least one track must have $p_T>1.0$~\GeV.
These requirements ensure that there is  sufficient information to
identify the leading jet as a heavy flavor candidate and have a typical efficiency
of about 90\%.  
To enrich the sample with heavy flavor jets,
a neural network based $b$-tagging algorithm ($b$-NN)~\cite{bidNIM} is used.
It exploits long decay lengths of $b$-flavored hadrons. 
The leading jet is required to pass a tight $b$-NN cut $>0.225$~\cite{bidNIM}.

Data events with a single \ppbar collision vertex (``\onevtx'' sample), which contain 
DP candidates, are selected separately from events with two vertices (``\twovtx'' sample), which
contain DI candidates.
The collision vertices in both samples
are required to have at least three associated tracks 
and to be within 60~cm of the center of the detector along the beam ($z$) axis. 
The total number of \gpTHRj and \gpHFjj candidate events, 
referred to below as inclusive and heavy flavor (HF) samples, 
in each of the \onevtx or \twovtx categories after all selection criteria 
have been applied are given in Table~\ref{tab:12vtx_data}.
No requirement on the origin vertex for the photon or 
jets is imposed here for the \twovtx events.

\begin{table}[htpb]
\begin{center}
\caption{The numbers of selected \onevtx and \twovtx candidate events, $N_{\rm 1vtx}$ and $N_{\rm 2vtx}$, 
and their ratio in the  \gpTHRj (inclusive) 
and  \gpHFjj (HF) samples.}
\label{tab:12vtx_data}
\begin{tabular}{cccc} \hline\hline
  Data       &    &  &  \\ 
  Sample     & $N_{\rm 1vtx}$ & $N_{\rm 2vtx}$ & $N_{\rm 2vtx}/N_{\rm 1vtx}$\\\hline
  Inclusive ~& 218686 ~& 269445 ~& 1.23 $\pm$ 0.01\\
  HF        ~& 5004   ~&  5811  ~& 1.16 $\pm$ 0.02\\\hline\hline
\end{tabular}
\end{center}
\end{table}
\section{Data, signal, and background event models}
\label{sec:Models}

This section gives an overview of the DP and DI models built using
data and MC samples, to estimate the number of DP and DI events in data,~\ndp and \ndi.
These models are also used to calculate selection efficiencies and 
geometric and kinematic acceptances for DP and DI events.

\subsection{Signal models}
\label{sec:SignalModels}

\begin{itemize}
\item \textbf{DP data event model} (\mixdp)\textbf{:} 

The DP signal event model exploits the fact that two parton-parton
scatterings can occur in the same \ppbar collision.
Therefore, an individual signal DP event 
is constructed by overlaying one event from an inclusive 
data sample of $\gamma+\!\geq$1 jet data events with
another event from a sample of 
inelastic non-diffractive events selected with a minimum bias trigger and
a requirement of at least one reconstructed jet (``MB'' sample)~\cite{D02010,JES_NIM}. 
Both input samples contain only events with a single \ppbar collision vertex.
The \pt values of the jets from the MB event are recalculated relative to the vertex of 
the \gpj event.
The resulting mixed event  
is required to satisfy the same selection criteria
as applied to \gpTHRj data events with a single \ppbar collision.
The \mixdp sample
provides independent parton scatterings with \gpj and dijet final states, 
by construction.
In particular, since the \gpj process is dominated by small parton momentum fractions ($x$),
the $x$ values in the dijet production process remaining after
the first parton interaction occurred is generally unaffected, 
i.e. the two interactions have negligible correlation in the momentum space.
The mixing procedure is shown schematically in Fig.~\ref{fig:dp_mix}.
The \mixdp events shown in Fig.~2(b)
comprise about 60\% of both inclusive and HF samples.
\begin{figure}[h]
\hspace*{-2mm} \includegraphics[scale=0.3]{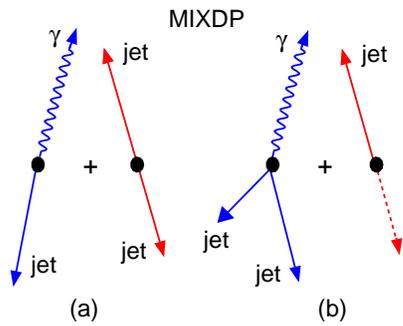}
~\\[-3mm]
\caption{(color online)
Schematic view of the mixing procedure 
used to prepare the \mixdp signal sample.
Two combinations are considered: 
(a) $\gamma+1$ jet and two jets from a dijet event and 
(b) $\gamma+2$ jets and one jet from a dijet event. 
The dotted line represents a jet failing the selection requirements
since this jet is either not reconstructed or beyond 
kinematic selection limits.}
\label{fig:dp_mix}
\end{figure}

\item \textbf{DI data event model} (\mixdi)\textbf{:}

The DI signal event model assures that the \gpTHRj DI events
originate from two separate $p\bar{p}$ collisions 
by preparing a mixture of $\gamma+\!\geq$1 jet events 
from the \gpj data and of MB events with requirements of $\geq$1 
selected jets and two \ppbar collision vertices for both data samples.
Thus, the second \ppbar collision contains only soft underlying energy
that can contribute energy to a jet cone, or a photon isolation cone.
In addition, in the case of $\geq$2 jets in either component of the \mixdi mixture
(i.e., in \gpj or MB events), 
the two leading jets are required to originate 
from the same vertex, using jet track information, 
as discussed in Appendix B of Ref.~\cite{D02010}.
Since the \pt of all reconstructed objects 
is calculated with respect to the primary $p\bar{p}$ collision vertex (PV0),
the jet \pt from the MB event is recalculated relative to the primary vertex of
the \gpj event (i.e., PV0 for the \twovtx data sample).
Here the PV0 is the $p\bar{p}$ collision vertex with the lowest
probability that it originates from a soft $p\bar{p}$ interaction~\cite{bidNIM}.
The resulting \gpTHRj events undergo the same selection as
applied to the data sample with two \ppbar collision vertices.

\end{itemize}

A fraction of the $\gamma+2$ jet events coming from one hard interaction
in the \mixdp and \mixdi models may be caused by DP events. 
This fraction was measured in Ref.~\cite{D02011}  as a function of the second jet \pt.
With our current selections, we have $\la p_T^{\rm jet2}\ra\approx 24$ GeV, 
and this fraction is expected to be around 4\%--5\%. Since in Eq.~\ref{eq:sig_eff}
we calculate the ratio of DP and DI events, and the fractions  of the $\gamma+2$ jet events
in the \mixdp and \mixdi models are similar,
it has been found that the corresponding DP fractions cancel.

To construct a model of \gpHFjj DP and DI signal events,
the leading jet in both the \mixdp and \mixdi samples 
should additionally satisfy the tight $b$ identification criteria
described in the previous section.

To create signal and background MC models for DP and DI events, 
we use an overlay of MC \gpj ($\gamma+b/c$-jet) and dijet events.
These events are generated with {\sc pythia} or
{\sc sherpa}~\cite{SHERPA} event generators and processed
through a {\sc geant}-based~\cite{GEANT} simulation of the D0 detector
response.  To accurately model the effects of multiple
\ppbar interactions and detector noise, data events from random
\ppbar crossings are overlaid on the MC events using data from the same data taking
period as considered in the analysis.
These MC events are then processed using the same reconstruction code as for data.
We also apply additional smearing to the reconstructed photon and jet \pt
so that the measurement resolutions in MC match those in data.
These MC events are used to create single- and two-vertex samples. 

\begin{itemize}

\item \textbf{DP and DI MC models} (\mcdp and \mcdi)\textbf{:}

Using the \gpj ($\gamma+b/c$-jet) and dijet MC samples, we create 
\gpTHRj (\gpHFjj) DP and DI MC models,
similar to those constructed for \mixdp and \mixdi data samples,
by examining information for jets and the photon
at both the reconstructed and particle level.
These samples are used to calculate efficiencies and acceptances for DP and DI events. 
As a cross check, we have compared
\pt and $\eta$ distributions for the jets and the photon at the reconstruction
level in these models with those
in the \mixdp and \mixdi data samples. Small discrepancies have been
resolved by reweighting the MC spectra and creating models denoted as data-like \mcdp and \mcdi.

\end{itemize}

\subsection{Background models}

To extract fractions of DP and DI events from data, 
we need to build SP background models.

\begin{itemize}

\item \textbf{SP one-vertex event model} (\sponevtx)\textbf{:}

A background to the DP events are single parton-parton scatters 
with two additional bremsstrahlung jets resulting in a \gpTHRj final state
in a single \ppbar collision event. 
To model this background, we consider a sample of MC \gpTHRj events
generated with MPI modeling removed.
The \sponevtx sample contains the final state with a photon, leading jet, and two additional
bremsstrahlung jets with the same selection criteria as applied to the data
sample with a single \ppbar collision vertex.
The {\sc sherpa} SP model is taken as the default.

\item \textbf{SP two-vertex event model} (\sptwovtx)\textbf{:}

The background to DI events differs from the \sponevtx model
in that the \gpTHRj MC events are selected with 
two reconstructed \ppbar collision vertices.
Events with no jet activity in the second vertex are selected by requiring
the three jets to originate from the primary \ppbar collision vertex.

\end{itemize}

To model the background to the \gpHFjj DP and DI processes, 
the \sponevtx and \sptwovtx samples are constructed using the same 
techniques, but using $\gamma+b/c$-jet events generated with the {\sc pythia} and {\sc sherpa} MCs 
with MPI modeling removed.

\section{Discriminating variable}
\label{sec:Vars}

\begin{figure}[htp!]
\hspace*{2mm} \includegraphics[scale=0.4]{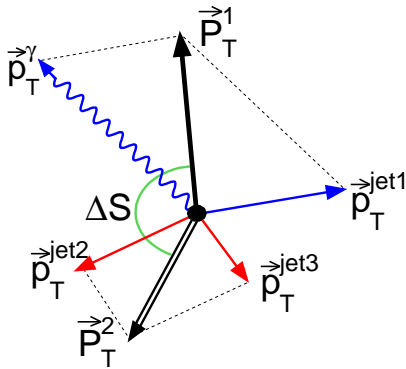}
~\\[-15mm]
\caption{(color online)
A possible configuration of photon and jets transverse momenta vectors in \gpTHRj events.
Vectors $\vec{P}_{\rm T}^1$ and $\vec{P}_{\rm T}^2$ are the \pt-imbalance vectors of \gpj and dijet pairs. }
\label{fig:deltaPHI}
\end{figure}

Unlike the SP scattering $2 \rightarrow 4$ process, which produces
a $\gamma$ + jet final state and two bremsstrahlung jets, 
the DP mechanism has two independent $2 \rightarrow 2$ 
parton-parton scatterings within the same \ppbar collision, resulting in
substantially different kinematic distributions in the final state.
Discrimination between these processes is obtained by examining the
azimuthal angle between the \pt vectors of two object pairs 
in \gpTHRj events,
\begin{eqnarray}
\Delta S \equiv \Delta\phi\left(\vec{P}_{\rm T}^1,~\vec{P}_{\rm T}^2\right),
\label{eq:deltaPHI}
\end{eqnarray}
where $\vec{P}_{\rm T}^1 = \vec{p}_{\rm T}^{~\gamma} + \vec{p}_{\rm T}^{\rm ~jet_1}$
and $\vec{P}_{\rm T}^2 = \vec{p}_{\rm T}^{\rm ~jet_2} + \vec{p}_{\rm T}^{\rm ~jet_3}$.
Figure~\ref{fig:deltaPHI} illustrates a possible orientation of photon and 
jets transverse momentum vectors in \gpTHRj events, as well as the vectors
$\vec{P}_{\rm T}^1$ and $\vec{P}_{\rm T}^2$.

The differential cross section as a function of \dPhi was measured in Ref.~\cite{D02011}
and compared with various SP and MPI models.
Momentum conservation causes \dPhi to peak near $\pi$, and this is
particularly visible in SP, although
detector resolution effects and additional gluon radiation produce a significant
number of events at smaller angles.
For DP events, where the photon and leading jet usually
come from one parton-parton scattering and the two other jets usually come from another
parton-parton scattering, the pairwise balance \dPhi angle has no pronounced peak at any
particular value, although some residual bias remains towards $\Delta S = \pi$ caused by the DP events shown in Fig.~2(b).

\section{Fractions of DP and DI events}
\label{sec:Frac}

\subsection{Fractions of DP events}
\label{sec:DPfraction}

To calculate \sigmaeff, we need the number of DP events (\ndp) 
in Eq.~(\ref{eq:sig_eff}), given by the product of the fraction of DP events (\fdp) and
the size of the \onevtx sample.  The fraction \fdp is estimated 
in the \gpTHRj \onevtx data sample using the DP (\mixdp) and SP (\sponevtx) models.
The DP fractions (and \sigmaeff) are measured 
in the inclusive and HF samples separately.

The fraction \fdp is found using a maximum likelihood fit~\cite{HMCMLL}
of the \dPhi distribution of the data to signal and background templates
that are taken to be the shapes of the \dPhi distribution
in the \mixdp and \sponevtx models, respectively.
Signal and background samples used as templates, described in Section \ref{sec:Models}, 
satisfy all the selection criteria applied to the data sample.

\begin{figure}[htb!]
\vspace*{-3mm}
\hspace*{4mm} \includegraphics[scale=0.34]{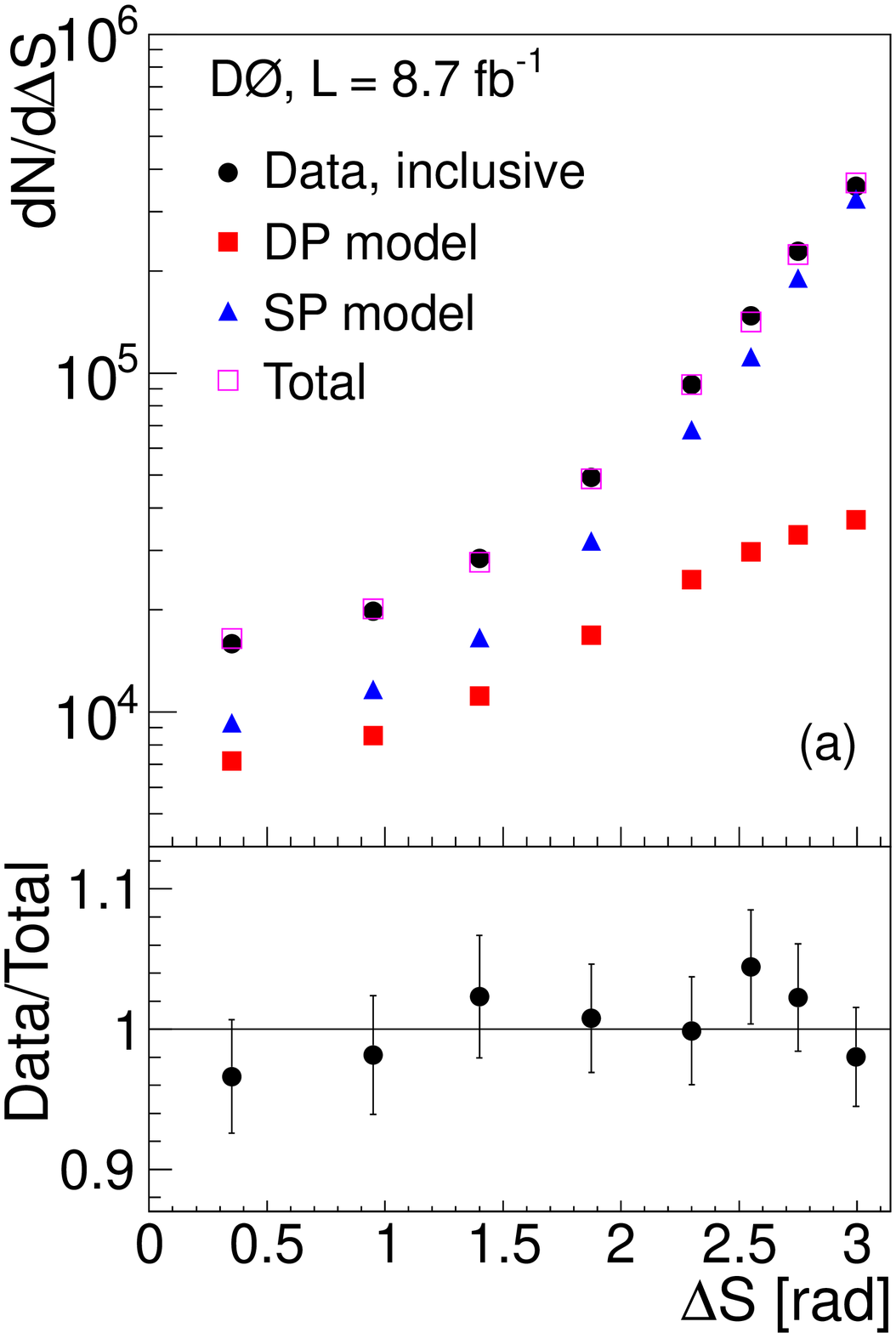}
\hspace*{4mm} \includegraphics[scale=0.34]{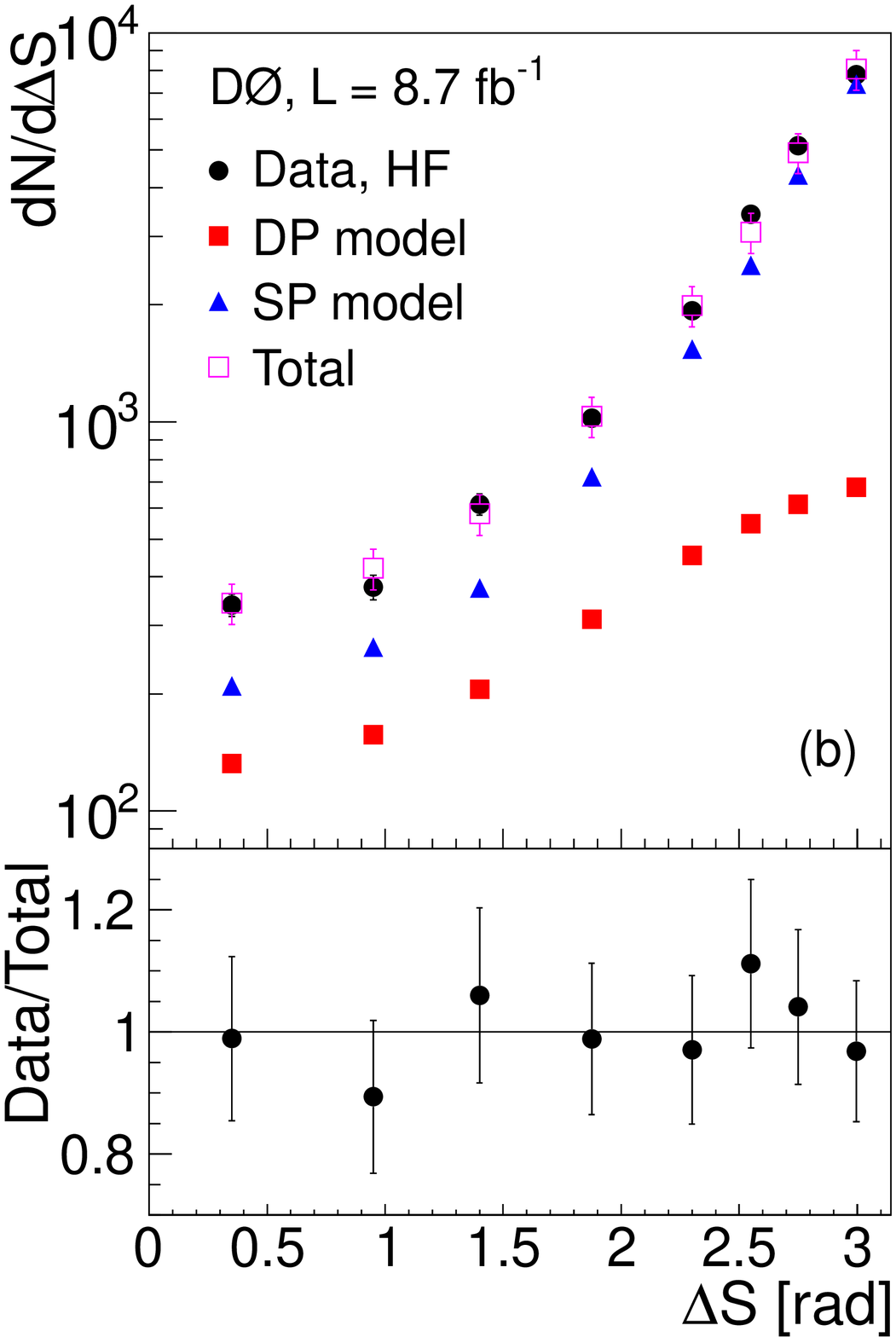}
~\\[-3mm]
\caption{(color online)
The \dPhi distribution in the data, DP and SP models,
and the sum of the DP and SP contributions 
weighted with their fractions (``Total''). 
The plots (a) and (b) correspond to the inclusive and HF samples, respectively.
The lower subplots
show the relative difference of the data points with respect to
the fitted sum, along with the total uncertainties, i.e.
DP fraction and statistical uncertainties from 
data and MC added in quadrature.
}
\label{fig:dpfit_init}
\end{figure}

A first approximation to the fractions can be obtained
from the fits to inclusive and HF data shown in Fig.~\ref{fig:dpfit_init}.  
The measured DP fractions are: 
\begin{equation}
f_{\rm DP}^{\rm inc} = 0.202 \pm 0.007 
\label{eq:fdp_inc_def}
\end{equation}
and
\begin{equation}
f_{\rm DP}^{\rm HF} = 0.171 \pm 0.020,
\label{eq:fdp_HF_def}
\end{equation}
respectively. 
If it is not stated otherwise, the uncertainties shown in 
Eqs.~\ref{eq:fdp_inc_def}, \ref{eq:fdp_HF_def}, and in the text below are only statistical.
The sum of DP and SP models weighted with their fractions
describes the data with \chindf = 0.45 for the inclusive case and \chindf = 0.26 
(with the number of degrees of freedom, $ndf$ = 7) for the HF sample,
i.e. $\approx 87\%$ and $\approx 97\%$ $\chi^2$-probability, respectively.

While the default SP model obtained with {\sc sherpa}, 
provides a reasonable description of the \dPhi distribution in data,
it might be not perfect for other related kinematic variables,
which may affect the DP fractions as well.
For this reason we examine two alternative models.
Since the fraction of events with the leading jet coming from the second parton
interaction is small ($\lesssim10\%$), the \DPhi distribution
(the azimuthal angle between the photon and leading jet \pt vectors)
in the inclusive \gpTHRj events should be sensitive to initial and final state radiation
effects in the \gpj events.
We construct a modified \gpTHRj SP model in which the MC \DPhi distribution
is reweighted to agree with data, as discussed in the Appendix.
The $f_{\rm DP}^{\rm inc, rew_1}$ fraction obtained with the \DPhi reweighted SP model is $0.216 \pm 0.007$.
The shapes of the \pt spectra of the second and third jets are important for the
\dPhi calculation.  To estimate the effects of possible mismodeling
of the jet \pt spectra, we create an alternative SP model by reweighting the jet
\pt distributions in the default MC SP model in two dimensions 
(\pt of the second and third jet) to SP data. 
After reweighting, the DP fraction is recalculated and found to
be $f_{\rm DP}^{\rm inc, rew_2} = 0.195 \pm 0.007$.
The sum of DP and the \DPhi (jet \pt) reweighted SP models weighted 
with their fractions describes the data with \chindf = 0.51 (\chindf = 0.43), $ndf$ = 7.

The fraction obtained by averaging \fdp values after reweighting the \DPhi and second and third
jet \pt spectra is used as a central value,
and the difference between this and the value
obtained with the default SP model (Eqs.~\ref{eq:fdp_inc_def} and~\ref{eq:fdp_HF_def}) 
is taken as a systematic uncertainty.
The final DP event fraction in the inclusive sample is 
\begin{equation}
f_{\rm DP}^{\rm inc,avg} = 0.206 \pm 0.007\thinspace({\rm stat}) \pm 0.004\thinspace({\rm syst}). 
\label{eq:fdp_inc_rew}
\end{equation}
A similar reweighting procedure and determination of central 
value and the assignment of uncertainties is applied for 
the SP model in the HF sample, and the DP fraction is found to be
\begin{equation}
f_{\rm DP}^{\rm HF,avg} = 0.173 \pm 0.020\thinspace({\rm stat}) \pm 0.002\thinspace({\rm syst}).
\label{eq:fdp_inc_rew}
\end{equation}
All the results on DP fractions are summarized in Table~\ref{tab:dp_fraction}.
\begin{table}[htpb!]
	\centering
	\small
	\caption{DP event fraction for different reweighting procedures.}
	\label{tab:dp_fraction}
	\begin{tabular}{ccc} \hline\hline
	\fdp ~&~ Inclusive sample ~&~ HF sample \\\hline
	No reweighting ~&~ $0.202 \pm 0.007$ ~&~ $0.171 \pm 0.020$ \\ 
	\DPhi reweighted ~&~ $0.216 \pm 0.007$ ~&~ $0.169 \pm 0.020$ \\ 
	$p_{T}^{\rm jet2}$ and $p_{T}^{\rm jet3}$ reweighted ~&~ $0.195 \pm 0.007$ ~&~ $0.177 \pm 0.020$ \\ 
	Reweighted average ~&~ $0.206 \pm 0.007$ ~&~ $0.173 \pm 0.020$ \\ \hline\hline
	\end{tabular}
\end{table}

The measured DP fraction is lower than that measured in the earlier D0 analysis
\cite{D02010}. This is primarily because of the smaller jet cone radius used in
the current measurement ($R=0.5$ vs. $R=0.7$ in \cite{D02010}), 
what leads to a smaller probability to pass the jet reconstruction threshold 
(6 \GeV for the uncorrected jet \pt). 
The use of a smaller jet cone also significantly reduces the dijet cross section (a factor of $1.5-2$) in
the \pt region of interest.
Because the second parton interaction produces mostly a dijet final state,
the measured DP fraction drops.

In addition to the SP events produced in single $p\bar{p}$ collisions,
another source of possible background to the single-vertex \gpTHRj DP events 
are two $p\bar{p}$ collisions produced very close to each other along the beam direction, 
so that a single vertex is reconstructed.
This contribution is estimated using the instantaneous luminosity, 
the bunch size, the time between bunch crossings, and the vertex resolution, 
and found to be negligible at a level of~$\lesssim0.2\%$.
%this background contributes $<10^{-3}$ of the final sample.

\subsection{Fractions of DI events}

In addition to \fdp, the fraction of DI events  (\fdi) occurring in events with two \ppbar
collisions within the same bunch crossing must be determined to measure $\sigma_{\rm eff}$.
A discriminant is constructed using the track information of a jet and 
of the assignment of tracks to the two \ppbar collision vertices (PV0 and PV1).
We use the \pt-weighted position along the beam
($z$) axis of all tracks associated to the jet and
the fraction of charged particles in the jet (CPF). 
The CPF discriminant is based on  the fraction of total charged particles' transverse momentum 
(i.e., total track \pt) in each jet $i$ originating from each identified vertex $j$ in the event:\\[-4mm]
\begin{eqnarray}
{\rm CPF(jet}_i,{\rm vtx}_j) = \frac{\sum_k p_T({\rm trk}_k^{{\rm jet}_i},{\rm vtx}_j)}{\sum_n\sum_l p_T({\rm trk}_l^{{\rm jet}_i},{\rm vtx}_n)}.
\end{eqnarray}
Each jet is required to have  CPF $\gt$ 0.5 and at least two tracks.

In events with two \ppbar collisions, jets in \gpTHRj events may originate
either from PV0 or PV1. 
The leading jet is required to originate from PV0.
Four classes of events are defined:
\begin{enumerate}[I:]
%\roman{enumi}
\item All three jets originate from PV0. \\[-6mm]
\item Jet 1 and jet 2 originate from PV0 while jet 3 originates from PV1.\\[-6mm]
\item Jet 1 and jet 3 originate from PV0 while jet 2 originates from PV1.\\[-6mm]
\item Jet 1 originates from PV0 while jet 2 and jet 3 originate from PV1.\\[-4mm]
\end{enumerate}

Class I corresponds to a type of \gpTHRj event that has all three jets originating 
from the same \ppbar collision with no reconstructed jets in the other, i.e., background (non-DI) events,
while classes II, III and IV correspond to three types of signal (DI) events.

To assign a jet to a vertex and extract \fdi using the jet track information,
we need the $z$ resolution of the jet-to-vertex assignment algorithm, $\sigma_z$.
This resolution can be calculated in the \gpTHRj data event sample with a single 
\ppbar collision.
Since these events have only one reconstructed \ppbar collision vertex, all the jets
should originate from this vertex.
To find the $z$ position of a jet's origin, 
we consider all tracks inside a jet cone
and calculate the \pt-weighted position in $z$ of all the tracks ($z_{\rm jet}$).
The track $z$ position is calculated at the point of closest approach of
each track to the beam axis.
For each jet in the \onevtx data sample,
we estimate the distance between the $z_{\rm jet}$ and 
the $z$-vertex position, $\Delta z({\rm vtx,~jet})$.
We find $\sigma_z \approx$~1.2~cm
and that 98\%--99\% of jets in \onevtx events have $\Delta z({\rm vtx,~jet})<3\sigma_z$.
We consider a jet to originate from a vertex if $|z - z_{\rm jet}| < 3\sigma_z$.
If the jet is located within $3\sigma_z$ of both vertices it is assigned to the closest vertex.

Table \ref{tab:di_classes} shows the fractions of \twovtx data events in each class.
From this table, one can see that
the single interaction events (Class I) dominate over DI events (sum of classes II, III, and IV). 
The DI event fraction is $f_{\rm DI} = 0.135 \pm 0.002$ for the inclusive sample
and $f_{\rm DI}^{\rm HF} = 0.131 \pm 0.010$ for the sample with a heavy flavor
leading jet.

\begin{table}[htpb]
\begin{center}
\caption{The fractions of \twovtx data events for Class I (non-DI events),
and three classes of DI events in the  \gpTHRj (inclusive) 
and  \gpHFjj (HF) samples.}
\label{tab:di_classes}
\begin{tabular}{ccc} \hline\hline
  DI event class ~& Inclusive sample ~& HF sample \\\hline
  I   ~&  0.865 $\pm$ 0.001 ~& 0.869 $\pm$ 0.010 \\
  II  ~&  0.074 $\pm$ 0.001 ~& 0.078 $\pm$ 0.008 \\
  III ~&  0.044 $\pm$ 0.001 ~& 0.040 $\pm$ 0.006 \\
  IV  ~&  0.017 $\pm$ 0.001 ~& 0.013 $\pm$ 0.003 \\\hline\hline
\end{tabular}
\end{center}
\end{table}

The distance in $z$ between two vertices $\Delta z({\rm PV0, PV1})$ 
may affect the measured DI fraction,
since about 5\% of events have $\Delta z({\rm PV0, PV1})<3\sigma_z$. 
No requirement is placed on this distance in the analysis.  To quantify the
dependence of the DI fraction on this distance, we have also measured the DI fraction with the
requirement that the two vertices are separated
by $\Delta z({\rm PV0, PV1})>5\sigma_z$.
Table~\ref{tab:dZjet} shows \fdi for the two cases: 
no cut and $\Delta z({\rm PV0, PV1})>5\sigma_z$.
The difference between them is taken as a systematic uncertainty.
\begin{table}[htpb!]
        \centering
        \small
        \caption{DI event fraction with respect to $\Delta z$({\rm PV0, PV1}).}
        \label{tab:dZjet}
        \begin{tabular}{ccc} \hline\hline
        $\Delta z$({\rm PV0, PV1}) ~&~ Inclusive sample ~&~ HF sample \\\hline
        All values ~&~ $0.135 \pm 0.002$ ~&~ $0.131 \pm 0.010$ \\
        $\gt 5\sigma_z$ ~&~ $0.129 \pm 0.002$ ~&~ $0.122 \pm 0.011$ \\\hline\hline
        \end{tabular}
\end{table}

An additional uncertainty is due to the determination of the photon vertex.
This uncertainty has been estimated using events with a photon EM cluster in the central
region ($|\eta^{\gamma}_{\rm det}|<1.0$) with a matched CPS cluster.
These events allow us to extrapolate the photon direction along the $z$~axis and determine the vertex 
position on the $z$~axis~\cite{EMID_NIM}.
Using the \gpTHRj data, we estimate the photon pointing resolution in $z$ to be about 4.5~cm.
Using this resolution and 
the distribution of the distance in $z$ between the first and second vertices in \twovtx events,
we find that the photon origin vertex may be 
misidentified in about 4\% of events, which is taken as a systematic uncertainty.

The DI fractions extracted for the inclusive and heavy flavor samples are:
\begin{equation}
\label{eq:fdi}
f_{\rm DI} = 0.135 \pm 0.002\thinspace({\rm stat}) \pm 0.008\thinspace({\rm syst}), 
\end{equation}
\begin{equation}
\label{eq:fdihf}
f_{\rm DI}^{\rm HF} = 0.131 \pm 0.010\thinspace({\rm stat}) \pm 0.011\thinspace({\rm syst}). 
\end{equation}

A cross check of the measured DI fractions is performed by fitting the \dPhi  templates for signal and
background models to data as was done to extract the DP fraction in
Section~\ref{sec:DPfraction}.
We use the \mixdi sample for the signal template and 
the \sptwovtx sample for the background template (see Section~\ref{sec:Models}).
The measured fractions $f_{\rm DI} = 0.127 \pm 0.021$ (with \sptwovtx model
taken from {\sc sherpa}) and $f_{\rm DI} = 0.124 \pm 0.056$ (\pythia)
are in good agreement with each other and with $f_{\rm DI}$ obtained by the jet-track
method.  The results for the heavy flavor jet sample are
$f_{\rm DI}^{\rm HF} = 0.153 \pm 0.044$ with the SP model from {\sc sherpa} 
and $f_{\rm DI}^{\rm HF} = 0.143 \pm 0.056$ 
using \pythia, which are also in agreement with the jet-track method.
Since the results of this cross check agree with the values obtained using the jet track method,
we do not assign an additional systematic uncertainty.

\section{DP and DI efficiencies, $\bm R_{\bm c}$ and $\bm \sigma_{\bf hard}$}
\label{sec:Eff}

\subsection{Ratio of signal fractions in DP and DI events}
\label{sec:sigfrac}
A fraction of events with jets containing energetic $\pi^0$ or $\eta$ mesons may
satisfy the photon selection criteria. 
The photon fraction in the selected data is estimated using
the maximum likelihood fit of templates from the output of the photon
identification neural network ($O_{\rm NN}$) in signal and background events to that in data,
as described in detail in Ref.~\cite{gamjet_tgj}.
The photon fractions in DP and DI events are found to be similar.
For example, for a photon in the central calorimeter (CC) region,
$f^{\gamma,CC}_{\rm DP} = 0.432 \pm 0.002$ and $f^{\gamma,CC}_{\rm DI} = 0.437\pm 0.004$
for DP and DI events, respectively. 
The photon fractions are slightly higher in the forward region due to tighter photon selections.

The fractions of events with $b$ or $c$~jets in the \onevtx and \twovtx data samples are estimated
using templates for the invariant mass of charged particle tracks associated with the secondary vertex,
$M_{\rm SV}$ (see Ref.~\cite{gamma_b}) for $\gamma+b/c$-jet and \gpj MC samples.
The resulting HF fractions are dominated by $c$~quarks,
$f^b_{\rm DP} = 0.352\pm 0.025, f^c_{\rm DP} = 0.551\pm 0.041$,
and $f^b_{\rm DI} = 0.327\pm 0.019, f^c_{\rm DI} = 0.573\pm 0.043$.
The HF fractions in DP and DI samples are in good agreement.
Approximately 10\% of the jets tagged as HF come from mistagged light quark jets.

The overall signal fractions in DP and DI samples
and their ratio in the inclusive, $f^\gamma_{\rm DP}/f^\gamma_{\rm DI}$, and
HF samples, $(f^\gamma_{\rm DP}f^{\rm HF}_{\rm DP})/(f^\gamma_{\rm DI}f^{\rm HF}_{\rm DI})$, 
are summarized in Table~\ref{tab:dpdi_frac}.
The systematic uncertainties on the signal fraction are
caused by the uncertainties on the photon and heavy flavor fractions from
$O_{\rm NN}$ and $M_{\rm SV}$ template fitting.

\begin{table}[htpb]
\begin{center}
\caption{The overall signal fractions in DP and DI samples and their ratio
in the \gpTHRj (inclusive) and \gpHFjj (HF) samples.
Total uncertainties are shown, i.e., statistical and systematic uncertainties added in quadrature.}
\label{tab:dpdi_frac}
\begin{tabular}{ccccc} \hline\hline
  Sample    ~& DP ~& DI ~& ratio \\\hline
  Inclusive ~& 0.445 $\pm$ 0.005   ~&  0.456 $\pm$ 0.008 ~& 0.976 $\pm$ 0.019\\
  HF        ~& 0.402 $\pm$ 0.030   ~&  0.405 $\pm$ 0.030 ~& 0.993 $\pm$ 0.104\\\hline\hline
\end{tabular}
\end{center}
\end{table}

\subsection{Ratio of signal efficiencies in DP and DI events}

The selection efficiencies for DP and DI events enter Eq.~(\ref{eq:sig_eff})
only as ratios, substantially canceling correlated systematic uncertainties.
The DP and DI events differ from each other by the number of $p\bar{p}$
collision vertices  (one vs.\ two), and therefore 
their selection efficiencies $\epsdi$ and $\epsdp$ may differ
due to different amounts of soft unclustered energy in the single and double
$p\bar{p}$ collision events.
This could lead to a difference in the jet reconstruction efficiencies
because of the different probabilities for jets to pass the
$p_T \gt 6$~GeV requirement applied during jet reconstruction.
It could also lead to different photon selection efficiencies because of
different amounts of energy in the track and calorimeter isolation cones
around the photon. To estimate these efficiencies, we use the data-like \mcdp and \mcdi samples
described in Section~\ref{sec:Models}.

Using these models, we find the ratio of the geometric and kinematic acceptances
for DP and DI events to be
$A_{\rm DP}/ A_{\rm DI} = 0.551 \pm 0.010\thinspace({\rm stat}) \pm 0.030\thinspace({\rm syst})$
for the inclusive sample and
$A_{\rm DP}^{\rm HF}/ A_{\rm DI}^{\rm HF} = 0.567 \pm 0.021\thinspace({\rm stat}) \pm 0.052\thinspace({\rm syst})$
for the HF sample.
The difference between the $A_{\rm DP}$ and $A_{\rm DI}$ acceptances 
is caused by an average difference of 0.5~GeV in jet \pt due to the
offset energy entering the jet cone from the second vertex~\cite{JES_NIM}.
This significantly increases
the reconstruction efficiency of jets (mainly for second and third jets) in DI events.
The differences between the acceptances obtained with data-like 
and default \mcdp and \mcdi models are taken as systematic uncertainty.
An additional systematic uncertainty (about 1\%) is caused by the difference between photon
identification efficiencies obtained with {\sc sherpa} and {\sc pythia}.
For the HF sample, we also correct for the $b$-tagging selection efficiency. 
The ratio of the HF jet selection efficiencies is 
$\varepsilon^{\rm HF}_{\rm DP}/\varepsilon^{\rm HF}_{\rm DI} = 1.085 \pm 0.019$.
This number is obtained by weighting $b$- and $c$-jet efficiencies
with their fractions found in Section~\ref{sec:sigfrac}.
The typical HF jet selection efficiency is 60\% (10\%) for the tight $b (c)$ jet selection.
Only about 0.5\% of the light jets are misidentified as heavy flavor jets~\cite{bidNIM, gamma_b}.
The $b$-tagging efficiency decreases with increasing number of \ppbar collision vertices
due to a larger hit density in the SMT detector and a decrease in the track reconstruction
efficiency.
This also explains the lower $N_{\rm 2vtx}/N_{\rm 1vtx}$ ratio for the HF sample
compared to the inclusive sample in Table~\ref{tab:12vtx_data}.

\subsection{Vertex efficiencies}

The vertex efficiency $\varepsilon_{\rm 1vtx}$ ($\varepsilon_{\rm 2vtx}$)
corrects for single (double) collision events that are lost
in the DP (DI) candidate sample because of the single (double) vertex requirements
($|z_{\rm vtx}| < 60~\text{cm}$ and $\geq 3$ tracks). 
The ratio $\varepsilon_{\rm 1vtx}/\varepsilon_{\rm 2vtx}$
is calculated from the data and found to be $1.05 \pm 0.01$.
The probability to miss a hard interaction event having at least one jet with
$p_T>15$~GeV due to a non-reconstructed vertex is $<0.5\%$ and is ignored.

We might also have an additional fake reconstructed vertex that passes the vertex requirement.
This probability is estimated using \gpj events
and  $\gamma~+ \ge 3$ jet events simulated in MC without
zero-bias events overlay, as these events should contain only one vertex.
We find that the probability to have a second (fake) vertex is $<0.1\%$ and is ignored.

\subsection{Calculating $\bm R_{\bm c}$, \boldmath{$\sigma_{\rm hard}$}, \boldmath{$N_{\rm 1coll}$} and \boldmath{$N_{\rm 2coll}$}}

We calculate the numbers of expected events with
one ($N_{\rm 1coll}$) and two ($N_{\rm 2coll}$) $p\bar{p}$ collisions
resulting in hard interactions following the procedure of Ref.~\cite{D02010},
which uses the hard $p\bar{p}$ interaction cross section 
$\sigma_{\rm hard} = 44.76 \pm  2.89 ~{\rm mb}$.\ 
The values of $N_{\rm 1coll}$ and $N_{\rm 2coll}$ are obtained from 
a Poisson distribution parametrized with the average number of hard interactions in each
bin of the instantaneous luminosity $L_{\rm inst}$ distribution, 
$\la n \ra = (L_{\rm inst} / f_{\rm cross} ) \sigma_{\rm hard}$,
where $f_{\rm cross}$ is the frequency of beam crossings
for the Tevatron~\cite{d0det}. 
Summing over all $L_{\rm inst}$ bins, weighted with their fractions,
we get $R_c = (1/2) (N_{\rm 1coll} /N_{\rm 2coll}) (\ve_{\rm 1vtx} / \ve_{\rm 2vtx}) = 0.45$.
This number is smaller by approximately a factor of two compared to that
for the data collected earlier as reported in Ref.~\cite{D02010}.
Since $R_c$ and $\sigma_{\rm hard}$ enter Eq.~\ref{eq:sig_eff} for $\sigma_{\rm eff}$
as a product, any increase of $\sigma_{\rm hard}$ leads to an increase of $\la n \ra$
and, as a consequence, to a decrease in $R_c$, and vice versa.
Due to this partial cancellation of uncertainties, although the measured 
value of $\sigma_{\rm hard}$ has a 6\% relative uncertainty, the product $R_c\sigma_{\rm hard}$
only has a 2.6\% uncertainty, $R_c\sigma_{\rm hard} = 18.92 \pm 0.49$ mb.

\section{Results}
\label{sec:Sigma_eff}

Using Eq.~\ref{eq:sig_eff}, we obtain the following effective cross sections:
\begin{equation}
\label{eq:s_eff}
\sigma_{\rm eff}^{\rm incl} = 12.7 \pm 0.2\thinspace(\rm stat) \pm 1.3\thinspace(\rm syst) ~mb,
\end{equation}
\begin{equation}
\label{eq:s_eff_hf}
\sigma_{\rm eff}^{\rm HF} = 14.6 \pm 0.6\thinspace(\rm stat) \pm 3.2\thinspace(\rm syst) ~mb.
\end{equation}
Within uncertainties, the effective cross section in the inclusive event sample
is consistent with that in the event sample with identified heavy flavor jets.

The main sources of systematic uncertainties are summarized in
Table~\ref{tab:syst}. They are caused by uncertainties in the DP and DI fractions,
the ratio of efficiencies and acceptances in DP and DI events
(``$\varepsilon_{\rm DP}/\varepsilon_{\rm DI}$''),
signal fractions (``sig. frac.''), the uncertainty
in the ratio of the number of hard interactions with single and 
double $p\bar{p}$ hard collisions times $\sigma_{\rm hard}$
(``$R_c \sigma_{\rm hard}$''),
and jet energy scale (``JES'').
The latter is obtained from the variation of JES uncertainties
up and down by one standard deviation for all three jets~\cite{JES_NIM}.

\begin{table*}[htpb]
\caption{The systematic uncertainties from 
measurement of DP ($f_{\rm DP}$) and DI ($f_{\rm DI}$) fractions,
the ratio of efficiencies and acceptances in DP and DI events
(``$\varepsilon_{\rm DP}/\varepsilon_{\rm DI}$''),
signal fractions (``sig. frac.''), the uncertainty
in the ratio of the number of hard interactions with single and
double $p\bar{p}$ hard collisions times $\sigma_{\rm hard}$
(``$R_c \sigma_{\rm hard}$''),
and jet energy scale (``JES''),
shown together with overall
systematic ($\delta_{\rm syst}$), statistical ($\delta_{\rm stat}$) 
and total $\delta_{\rm total}$ uncertainties (in \%) for the $\sigma_{\rm eff}$ measurement. 
The total uncertainty $\delta_{\rm total}$ is calculated by adding the systematic and statistical uncertainties in quadrature.}
        \label{tab:syst}
        \begin{tabular}{cccccccccc} \hline\hline
 Data     & \multicolumn{6}{c}{Sources of systematic uncertainty} &   &  &  \\
 Sample   &  ~$f_{\rm DP}$~ & ~$f_{\rm DI}$~ & $\varepsilon_{\rm DP}/\varepsilon_{\rm DI}$ &~sig. frac.~&~ $R_c \sigma_{\rm hard}$ ~&~ JES ~&~$\delta_{\rm syst}$  & $\delta_{\rm stat}$ &  $\delta_{\rm total}$  \\\hline
 Inclusive &   3.9 &   6.5 &   5.6 &   2.0 &  2.6 &   2.9 &    10.4 &     1.8 &    10.6 \\
        HF &  11.6 &  11.2 &   9.4 &  10.4 &  2.6 &   1.3 &    21.6 &     4.0 &    22.0 \\\hline\hline
        \end{tabular}
\end{table*}

\begin{figure}[htp!]
\hspace*{2mm} \includegraphics[scale=0.45]{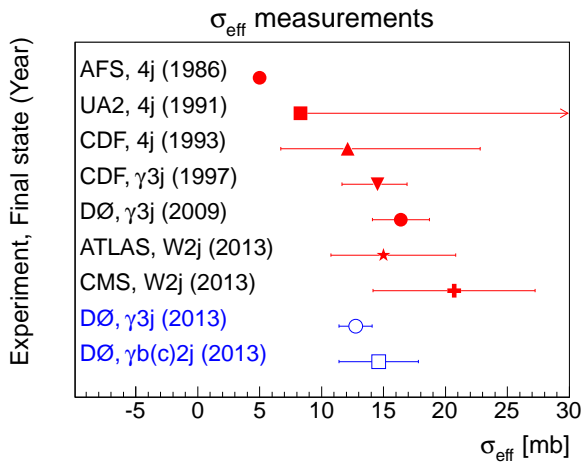}
\caption{(color online)
Existing  measurements of effective cross section, $\sigma_{\rm eff}$, 
compared with result presented here
(AFS: no uncertainty is reported; UA2: only a lower limit is provided).
}
\label{fig:sigmaEff_world}
\end{figure}

Figure~\ref{fig:sigmaEff_world} shows all existing measurements of \sigmaeff.
The $\sigma_{\rm eff}^{\rm incl}$ and \sigmaeffhf ~from this measurement agree both with the
previous D0~measurement~\cite{D02010}
and with those obtained by other experiments.
These new measurements of \sigmaeff ~are the most accurate to date, and 
also provide the first measurement involving heavy quarks.

\section{Discussion of PDF effects}
\label{sec:PDF}
The experimentally measured effective cross section $\sigma_{\rm eff}$,
presented in Eqs.~\ref{eq:s_eff} and~\ref{eq:s_eff_hf}, should be corrected
for the effect of double parton PDF (dPDF) evolution~\cite{Snigir2,Stirling,Snigir3}.
The dPDF evolution starts at a small scale $Q_0$, $\cal O$(1 GeV), where
the two PDFs corresponding to partons participating in DP scattering can be factorized. 
The dPDF evolution results in a correlation term at a larger energy scale $Q$, 
which necessitates the following correction: 
$[\sigma_{\rm eff}]^{-1} = [\sigma_{\rm eff}^{0}]^{-1} (1+\Delta(Q))$~\cite{Snigir2},
where $\Delta(Q)$ is a contribution induced by the dPDF correlation term, and
$\sigma_{\rm eff}^{0}$ depends only on the 
spatial distribution of parton flavors.
To estimate this correction factor, we have employed software,
provided by the authors of Ref.~\cite{Stirling}. 
It uses a numerical integration of the leading order DGLAP~\cite{DGLAP}
equation for the dPDFs, and which may be used to evolve the input dPDFs to any other scale.
To get access to the kinematics of the first and second parton interactions, the relevant part
of the {\sc pythia} code was modified for us by the {\sc pythia} authors. 
The evolution effect has been evaluated by examining the ratio
\begin{equation}
\label{eq:pdf_evol}
R_p(x_1,x_2;Q) = \frac{D_p(x_1,x_2;Q)}{D_p(x_1;Q)D_p(x_2;Q)}
\end{equation}
where $D_p(x_1,x_2;Q)$ is the dPDF with the parton momentum fractions $x_1, x_2$
of the two partons participating in the first and second parton interactions on the proton side 
at scale $Q$, and $D_p(x_{1(2)};Q)$ is a single parton MSTW2008LO PDF~\cite{MSTW}.
A similar equation can be written for the partons on the antiproton side.
Using the simulated \gpTHRj and \gpHFjj events, and applying our kinematic cuts, we have found 
the product of the two ratios $R_p R_{\bar{p}} = 1.01$ for \gpTHRj and 1.02 for \gpHFjj events.
This correction is expected to have a larger deviation from unity for higher $Q$
(e.g., it was found to be 0.93 for $\gamma$~+~${\rm 3~jet}$ 
at $\Ptg=70$ GeV that corresponds to the previous D0 measurement \cite{D02010}). 
In general, it should be calculated for each set of final states and kinematic selections.
Currently, the dPDF evolution implemented in Ref.~\cite{Stirling} is available at leading order accuracy, 
while having it at next-to-leading order would be preferable.
Due to the smallness of the found correction (1.01--1.02), and uncertainties 
related with the leading order approximation, this correction is 
not applied to the measured $\sigma_{\rm eff}$.

\section{Summary}
\label{sec:summary}
We have analyzed samples of \gpTHRj and \gpHFjj events
collected by the D0 experiment with an integrated luminosity of about 8.7~fb$^{-1}$
and determined the fractions of events with hard double parton
scattering occurring in a single $p\bar{p}$ collision at $\sqrt{s}=1.96$~TeV.
In the kinematic region $\Ptg > 26$~GeV,
$p^{\rm jet1}_{T} > $ 15~GeV, 15 $< p^{\rm jet2,3}_{T} < $ 35~GeV,
we observe that about (21 $\pm$ 1)\% and (17$\pm$ 2)\% of the events are produced in 
double parton interactions in the \gpTHRj and \gpHFjj final states.
The effective cross section \sigmaeff, which characterizes the spatial transverse parton
distribution in a nucleon, is found to be
$\sigma_{\rm eff}^{\rm incl} = 12.7 \pm 0.2\thinspace({\rm stat}) \pm 1.3\thinspace({\rm syst})$ mb
in \gpTHRj
and
$\sigma_{\rm eff}^{\rm HF} = 14.6 \pm 0.6\thinspace({\rm stat}) \pm 3.2\thinspace({\rm syst})$ mb
in \gpHFjj final states.

Our value of
\sigmaeff ~is in agreement with the results of previous
measurements and has a higher precision. 
This is the first measurement of \sigmaeff ~with heavy flavor jets in the final state. 
Due to the significant dominance of the Compton-like process (see Fig.~\ref{fig:qg_frac}),
we may conclude that there is no evidence for a dependence of $\sigma_{\rm eff}$ on the initial parton flavor.

~\\[5mm]
\centerline{\bf Acknowledgements}
~\\[1mm]

\input{acknowledgement}
\section{Appendix}
\label{Sec:App}

In Section \ref{sec:DPfraction}, we estimate the DP event fraction
using the predictions of Monte Carlo SP models.
In this appendix, we test variables that characterize 
the SP model and are related to the \dPhi distribution used to calculate
the DP fractions in Section \ref{sec:DPfraction}.

The variable \DPhi is sensitive to initial and final state radiation
and is strongly correlated to the \pt sum vector of the photon and leading
jet system, $\vec{P}_{\rm T}^1 = \vec{p}_{\rm T}^{~\gamma} + \vec{p}_{\rm T}^{\rm ~jet_1}$ (see Eq.~\ref{eq:deltaPHI}). 
We compare the distribution of \DPhi in the MC SP sample to data.
The latter is obtained after subtracting the DP contribution, predicted by the
DP data model \mixdp, according to the DP fractions in Eqs.~\ref{eq:fdp_inc_def} and~\ref{eq:fdp_HF_def}.

\begin{figure}[h]
\includegraphics[scale=0.34]{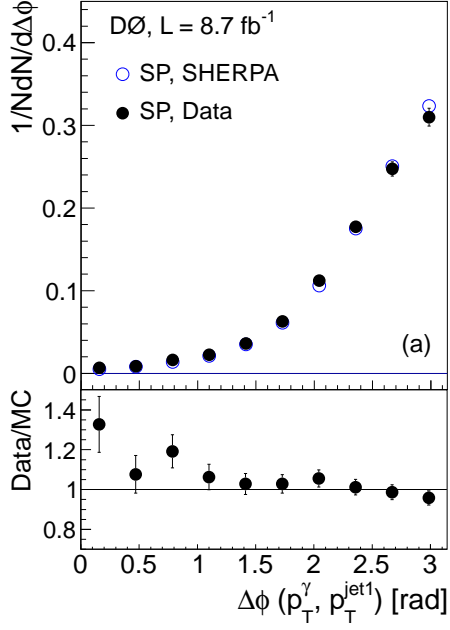}
\includegraphics[scale=0.34]{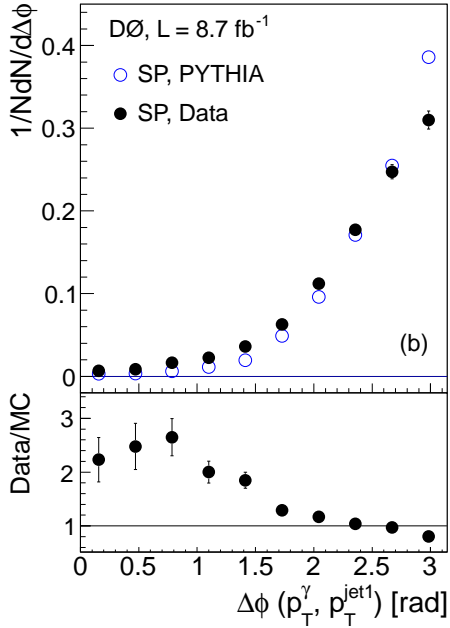}
\caption{(color online)
The \DPhi distribution in the SP model extracted from data
compared to that in (a) {\sc sherpa}, (b) {\sc pythia}.
The uncertainties shown are statistical only.}
\label{fig:dphigamjet1}
\end{figure}
\begin{figure}[h]
\includegraphics[scale=0.34]{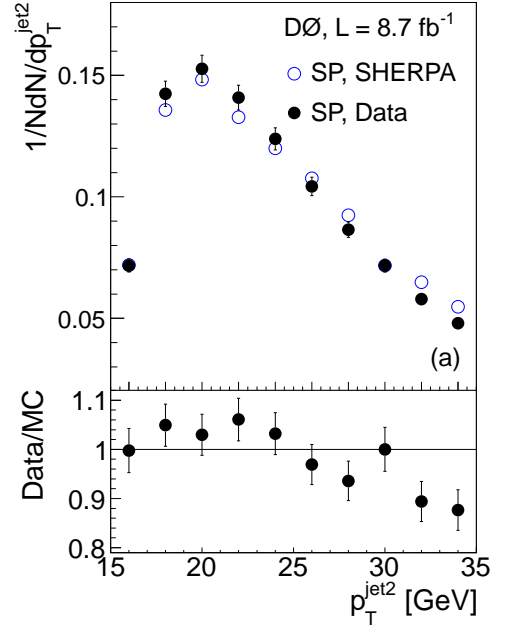}
\includegraphics[scale=0.34]{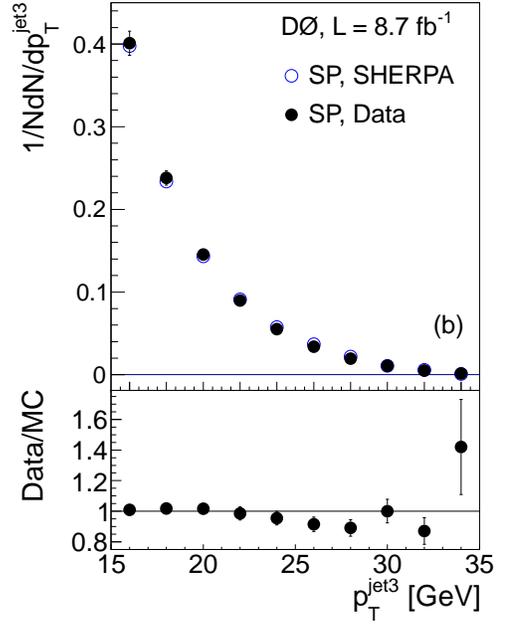}
\caption{(color online)
Spectra of the transverse momenta of (a) second and (b) third jets in
the \sherpa and data SP models.
The uncertainties shown are statistical only.}
\label{fig:jet2jet3}
\end{figure}

The comparison of the \DPhi spectra for the SP model extracted from data
with those in the {\sc sherpa} and {\sc pythia} MC generators is shown in
Fig.~\ref{fig:dphigamjet1}.
The {\sc sherpa} SP event model 
agrees better with the data  
compared to \pythia, where the \DPhi distribution is shifted towards $\pi$, resulting 
in much worse agreement with data. 
For this reason, the subsequent analysis is performed using the {\sc sherpa} SP model only.

The MC SP predictions for the \pt spectra of the second and third jets
are also important since, in addition to the vector $\vec{P}_{\rm T}^1$, 
they form the other imbalance vector of the \dPhi variable,
$\vec{P}_{\rm T}^2 = \vec{p}_{\rm T}^{\rm ~jet_2} + \vec{p}_{\rm T}^{\rm ~jet_3}$ (see Eq.~\ref{eq:deltaPHI}).
Figure~\ref{fig:jet2jet3} illustrates the transverse momenta of the
second and third jets of the {\sc sherpa} and data SP models.
Both jet-\pt spectra in {\sc sherpa} agree well with those in data.

However to construct a better (data-like) SP model,
the original default SP model from {\sc sherpa} is reweighted 
either in \DPhi bins, or in two dimensions of second and third jet \pt.
These two alternative data-like SP models are considered in Section~\ref{sec:DPfraction}
to calculate the DP fractions.
The later are compared to the DP fractions obtained with the default SP model
to derive related systematic uncertainties.

\bibliography{prd}
\bibliographystyle{apsrev}

\end{document}

%% file: author_list.tex
\affiliation{LAFEX, Centro Brasileiro de Pesquisas F\'{i}sicas, Rio de Janeiro, Brazil}
\affiliation{Universidade do Estado do Rio de Janeiro, Rio de Janeiro, Brazil}
\affiliation{Universidade Federal do ABC, Santo Andr\'e, Brazil}
\affiliation{University of Science and Technology of China, Hefei, People's Republic of China}
\affiliation{Universidad de los Andes, Bogot\'a, Colombia}
\affiliation{Charles University, Faculty of Mathematics and Physics, Center for Particle Physics, Prague, Czech Republic}
\affiliation{Czech Technical University in Prague, Prague, Czech Republic}
\affiliation{Institute of Physics, Academy of Sciences of the Czech Republic, Prague, Czech Republic}
\affiliation{Universidad San Francisco de Quito, Quito, Ecuador}
\affiliation{LPC, Universit\'e Blaise Pascal, CNRS/IN2P3, Clermont, France}
\affiliation{LPSC, Universit\'e Joseph Fourier Grenoble 1, CNRS/IN2P3, Institut National Polytechnique de Grenoble, Grenoble, France}
\affiliation{CPPM, Aix-Marseille Universit\'e, CNRS/IN2P3, Marseille, France}
\affiliation{LAL, Universit\'e Paris-Sud, CNRS/IN2P3, Orsay, France}
\affiliation{LPNHE, Universit\'es Paris VI and VII, CNRS/IN2P3, Paris, France}
\affiliation{CEA, Irfu, SPP, Saclay, France}
\affiliation{IPHC, Universit\'e de Strasbourg, CNRS/IN2P3, Strasbourg, France}
\affiliation{IPNL, Universit\'e Lyon 1, CNRS/IN2P3, Villeurbanne, France and Universit\'e de Lyon, Lyon, France}
\affiliation{III. Physikalisches Institut A, RWTH Aachen University, Aachen, Germany}
\affiliation{Physikalisches Institut, Universit\"at Freiburg, Freiburg, Germany}
\affiliation{II. Physikalisches Institut, Georg-August-Universit\"at G\"ottingen, G\"ottingen, Germany}
\affiliation{Institut f\"ur Physik, Universit\"at Mainz, Mainz, Germany}
\affiliation{Ludwig-Maximilians-Universit\"at M\"unchen, M\"unchen, Germany}
\affiliation{Panjab University, Chandigarh, India}
\affiliation{Delhi University, Delhi, India}
\affiliation{Tata Institute of Fundamental Research, Mumbai, India}
\affiliation{University College Dublin, Dublin, Ireland}
\affiliation{Korea Detector Laboratory, Korea University, Seoul, Korea}
\affiliation{CINVESTAV, Mexico City, Mexico}
\affiliation{Nikhef, Science Park, Amsterdam, the Netherlands}
\affiliation{Radboud University Nijmegen, Nijmegen, the Netherlands}
\affiliation{Joint Institute for Nuclear Research, Dubna, Russia}
\affiliation{Institute for Theoretical and Experimental Physics, Moscow, Russia}
\affiliation{Moscow State University, Moscow, Russia}
\affiliation{Institute for High Energy Physics, Protvino, Russia}
\affiliation{Petersburg Nuclear Physics Institute, St. Petersburg, Russia}
\affiliation{Instituci\'{o} Catalana de Recerca i Estudis Avan\c{c}ats (ICREA) and Institut de F\'{i}sica d'Altes Energies (IFAE), Barcelona, Spain}
\affiliation{Uppsala University, Uppsala, Sweden}
\affiliation{Taras Shevchenko National University of Kyiv, Kiev, Ukraine}
\affiliation{Lancaster University, Lancaster LA1 4YB, United Kingdom}
\affiliation{Imperial College London, London SW7 2AZ, United Kingdom}
\affiliation{The University of Manchester, Manchester M13 9PL, United Kingdom}
\affiliation{University of Arizona, Tucson, Arizona 85721, USA}
\affiliation{University of California Riverside, Riverside, California 92521, USA}
\affiliation{Florida State University, Tallahassee, Florida 32306, USA}
\affiliation{Fermi National Accelerator Laboratory, Batavia, Illinois 60510, USA}
\affiliation{University of Illinois at Chicago, Chicago, Illinois 60607, USA}
\affiliation{Northern Illinois University, DeKalb, Illinois 60115, USA}
\affiliation{Northwestern University, Evanston, Illinois 60208, USA}
\affiliation{Indiana University, Bloomington, Indiana 47405, USA}
\affiliation{Purdue University Calumet, Hammond, Indiana 46323, USA}
\affiliation{University of Notre Dame, Notre Dame, Indiana 46556, USA}
\affiliation{Iowa State University, Ames, Iowa 50011, USA}
\affiliation{University of Kansas, Lawrence, Kansas 66045, USA}
\affiliation{Louisiana Tech University, Ruston, Louisiana 71272, USA}
\affiliation{Northeastern University, Boston, Massachusetts 02115, USA}
\affiliation{University of Michigan, Ann Arbor, Michigan 48109, USA}
\affiliation{Michigan State University, East Lansing, Michigan 48824, USA}
\affiliation{University of Mississippi, University, Mississippi 38677, USA}
\affiliation{University of Nebraska, Lincoln, Nebraska 68588, USA}
\affiliation{Rutgers University, Piscataway, New Jersey 08855, USA}
\affiliation{Princeton University, Princeton, New Jersey 08544, USA}
\affiliation{State University of New York, Buffalo, New York 14260, USA}
\affiliation{University of Rochester, Rochester, New York 14627, USA}
\affiliation{State University of New York, Stony Brook, New York 11794, USA}
\affiliation{Brookhaven National Laboratory, Upton, New York 11973, USA}
\affiliation{Langston University, Langston, Oklahoma 73050, USA}
\affiliation{University of Oklahoma, Norman, Oklahoma 73019, USA}
\affiliation{Oklahoma State University, Stillwater, Oklahoma 74078, USA}
\affiliation{Brown University, Providence, Rhode Island 02912, USA}
\affiliation{University of Texas, Arlington, Texas 76019, USA}
\affiliation{Southern Methodist University, Dallas, Texas 75275, USA}
\affiliation{Rice University, Houston, Texas 77005, USA}
\affiliation{University of Virginia, Charlottesville, Virginia 22904, USA}
\affiliation{University of Washington, Seattle, Washington 98195, USA}
\author{V.M.~Abazov} \affiliation{Joint Institute for Nuclear Research, Dubna, Russia}
\author{B.~Abbott} \affiliation{University of Oklahoma, Norman, Oklahoma 73019, USA}
\author{B.S.~Acharya} \affiliation{Tata Institute of Fundamental Research, Mumbai, India}
\author{M.~Adams} \affiliation{University of Illinois at Chicago, Chicago, Illinois 60607, USA}
\author{T.~Adams} \affiliation{Florida State University, Tallahassee, Florida 32306, USA}
\author{J.P.~Agnew} \affiliation{The University of Manchester, Manchester M13 9PL, United Kingdom}
\author{G.D.~Alexeev} \affiliation{Joint Institute for Nuclear Research, Dubna, Russia}
\author{G.~Alkhazov} \affiliation{Petersburg Nuclear Physics Institute, St. Petersburg, Russia}
\author{A.~Alton$^{a}$} \affiliation{University of Michigan, Ann Arbor, Michigan 48109, USA}
\author{A.~Askew} \affiliation{Florida State University, Tallahassee, Florida 32306, USA}
% ...
%
% visitor_addresses.tex                       28 December 2013
%  available symbols are:
%  $\ast, \dag, \ddag, \S, \P, $\|$, $\ast\ast$, \dag\dag, \ddag\ddag ,\#
%
\collaboration{The D0 Collaboration\footnote{with visitors from
%{alton}
$^{a}$Augustana College, Sioux Falls, SD, USA,
%{burdin}
$^{b}$The University of Liverpool, Liverpool, UK,
%{grohsjean}
$^{c}$DESY, Hamburg, Germany,
%{de la cruz-burelo}
$^{d}$Universidad Michoacana de San Nicolas de Hidalgo, Morelia, Mexico
%{partridge}
$^{e}$SLAC, Menlo Park, CA, USA,
%{hesketh}
$^{f}$University College London, London, UK,
%{luna-garcia}
$^{g}$Centro de Investigacion en Computacion - IPN, Mexico City, Mexico,
%{santos}
$^{h}$Universidade Estadual Paulista, S\~ao Paulo, Brazil,
%{meyer}
$^{i}$Karlsruher Institut f\"ur Technologie (KIT) - Steinbuch Centre for Computing (SCC),
D-76128 Karlsrue, Germany,
%{patwa}
$^{j}$Office of Science, U.S. Department of Energy, Washington, D.C. 20585, USA,
%{cooke}
$^{k}$American Association for the Advancement of Science, Washington, D.C. 20005, USA
and
%{borysova}
$^{l}$Kiev Institute for Nuclear Research, Kiev, Ukraine
%{montgomery}
%$^{?}$Thomas Jefferson National Accelerator Facility, Newport News, VA 23606, USA,
%{falkowski}
%$^{?}$Laboratoire de Physique Theorique, Orsay, FR,
%{hooper,kozminski}
%$^{?}$}Visitor from Lewis University, Romeoville, IL, USA.
%{weber}
%$^{?}$Universit{\"a}t Bern, Bern, Switzerland.
%{deceased}
%{zanabria}
%$^{?}$City Colleges of Chicago, Chicago, IL, USA}
%$^{\ddag}$Deceased.
}} \noaffiliation
\vskip 0.25cm

%% file: acknowledgement.tex
% acknowledgement.tex                            28 December 2013
%
We are grateful to J.~R.~Gaunt, W.~J.~Stirling, T.~Sj\"ostrand and P.~Z.~Skands
for providing their codes and many helpful discussions.

We thank the staffs at Fermilab and collaborating institutions,
and acknowledge support from the
DOE and NSF (USA);
CEA and CNRS/IN2P3 (France);
MON, NRC KI and RFBR (Russia);
CNPq, FAPERJ, FAPESP and FUNDUNESP (Brazil);
DAE and DST (India);
Colciencias (Colombia);
CONACyT (Mexico);
NRF (Korea);
FOM (The Netherlands);
STFC and the Royal Society (United Kingdom);
MSMT and GACR (Czech Republic);
BMBF and DFG (Germany);
SFI (Ireland);
The Swedish Research Council (Sweden);
and
CAS and CNSF (China).

%% file: prd.bbl
\begin{thebibliography}{50}
\bibitem{d0_coordinate}
The polar angle $\theta$ and the azimuthal angle $\phi$ are defined
with respect to the positive $z$ axis, which is along the proton beam direction.
Pseudorapidity is defined as $\eta=-\ln[\tan(\theta/2)]$.
$\eta_{\rm det}$ and $\phi_{\rm det}$ are the pseudorapidity and the azimuthal angle
measured with respect to the center of the detector.
\end{thebibliography}
